\documentstyle[emulateapj,psfig]{article}

\begin{document}
\submitted{Submitted to ApJ Letters}

\title{The First Supernova Explosions in the Universe}

\author{Volker Bromm, Naoki Yoshida\altaffilmark{1},
and Lars Hernquist}

\affil{Harvard-Smithsonian Center for Astrophysics,
60 Garden Street, Cambridge, MA 02138;\\
vbromm@cfa.harvard.edu, nyoshida@cfa.harvard.edu, lars@cfa.harvard.edu}

\altaffiltext{1}{National Astronomical 
Observatory of Japan, Mitaka, Tokyo 181-8588, Japan},

\begin{abstract}
We investigate the supernova explosions that end the lives
of massive Population~III stars
in low-mass minihalos ($M\sim 10^{6}M_{\sun}$) at redshifts $z\sim 20$.
Employing the smoothed particle hydrodynamics method,
we carry out numerical simulations in a cosmological set-up
of pair-instability supernovae with explosion
energies of $E_{\rm SN}= 10^{51}$ and $10^{53}$ ergs. We find that
the more energetic explosion leads to the complete disruption of the
gas in the minihalo, whereas the lower explosion energy leaves much
of the halo intact. The higher energy supernova expels $\ga 90$\% of the
stellar metals into a region $\sim 1$~kpc across over a timescale of
$3-5$~Myr. Due to this burst-like initial star formation episode, a large
fraction of the universe could have been endowed with a metallicity
floor, $Z_{\rm min}\ga 10^{-4}Z_{\sun}$, already at $z\ga 15$.
\end{abstract}
\keywords{cosmology: theory --- galaxies: formation --- 
hydrodynamics --- intergalactic medium --- stars: formation ---
supernovae: general}

\section{INTRODUCTION}
One of the most important challenges in modern cosmology is to 
understand how the cosmic dark ages ended (e.g., Barkana \& Loeb 2001;
Bromm \& Loeb 2003a). Recent numerical simulations have presented
evidence that the first (Pop~III) stars, formed out of metal-free
gas in low-mass halos at redshifts $z\ga 20$, were predominantly
very massive with $M_{\ast}\ga 100 M_{\sun}$ (Bromm, Coppi, \& Larson 1999, 
2002; Abel, Bryan, \& Norman 2000, 2002). As the first stars mark the
crucial transition from a smooth, homogeneous universe to an increasingly
complex and structured one, the important question arises: {\it How did the
first stars die?}

The answer to this question sensitively depends on the precise mass of
a Pop~III star. In particular, if the star has a mass in the narrow
interval $140 \la M_{\ast} \la 260 M_{\sun}$, it will explode as a
pair-instability supernova (PISN), leading to the complete disruption
of the progenitor (Fryer, Woosley, \& Heger 2001; Heger et al. 2003).
Pop~III stars with masses below or above the PISN range are predicted
to form black holes. This latter fate is not accompanied by a significant
dispersal of heavy elements into the intergalactic medium (IGM), since
most of the newly synthesized metals will be locked up in the black
hole. The PISN, however, will contribute {\it all} its heavy element 
production to the surrounding gas.

How did the pristine IGM get enriched with the first heavy elements?
To account for the widespread presence of metals in the Ly$\alpha$
forest at $z\sim 5$ (e.g., Songaila 2001), star formation in low-mass
systems at $z\ga 10$ has been proposed as a likely source (Madau,
Ferrara, \& Rees 2001). The metals produced in these low-mass
halos can more easily escape from their shallow potential wells
than those at lower redshift (e.g., Aguirre et al. 2001a, b).
In addition,
the enriched gas has to travel much shorter distances between neighboring
halos at these early times, and it might therefore have been easier to
establish a uniform metal distribution in the IGM.

In this {\it Letter}, we present numerical simulations of the supernova
explosions that end the lives of massive Pop~III stars. In contrast to
earlier work, both analytical (e.g., Larson 1974; Dekel \& Silk 1986; Scannapieco,
Ferrrara, \& Madau 2002; Furlanetto \& Loeb 2003)
and numerical (e.g., Mori, Ferrara, \& Madau 2002; Thacker, Scannapieco, \&
Davis 2002;
Wada \& Venkatesan 2003), we here focus on the minihalos that are the
sites for the formation of the very first stars. The gas in these halos,
with masses of $M\sim 10^5 - 10^6 M_{\sun}$ and virializing at $z\ga 20$,
can only form stars in the presence of a sufficient amount of H$_{2}$,
the only viable coolant in these systems (e.g., Haiman, Thoul, \& Loeb 1996;
Tegmark et al. 1997; Yoshida et al. 2003a).

An important motivation for this numerical study is to substantiate
our earlier semi-analytical discussion of a fundamental transition
in the character of star formation at high redshifts, from very massive
Pop~III stars to lower-mass Pop~II stars at $z\ga 15$ 
(Mackey, Bromm, \& Hernquist 2003; see also Schneider et al. 2002).
This transition redshift is an important ingredient in determining
the reionization history of the universe, which has very recently
been constrained by the {\it WMAP} satellite (e.g., Cen 2003; Ciardi,
Ferrara, White 2003; Haiman \& Holder 2003; Wyithe \& Loeb 2003a,b;
Sokasian et al. 2003; Yoshida et al. 2003b).

\section{Numerical Methodology}

The evolution of the dark matter and gas components is calculated with
the parallel version of GADGET (Springel, Yoshida, \& White 2001) in
its ``conservative entropy'' formulation (Springel \& Hernquist 2002).
This code combines the smoothed particle hydrodynamics (SPH) method
with a hierarchical (tree) gravity solver. Here, we briefly describe
the additions to GADGET which are necessary for the investigation of
explosions in zero-metallicity gas.

We include all the relevant cooling processes for gas temperatures
$10^{2} \la T \la 10^{9}$K (Cen 1992; Bromm et al. 2002), as well as
heating due to the photoionization of H, He, and He$^{+}$.
The chemical reaction network comprises the 9 species H, H$^{+}$, H$^{-}$,
H$_{2}$, H$_{2}^{+}$, $e^{-}$, He, He$^{+}$, and He$^{++}$, including the
reactions given in Haiman et al. (1996). Our implicit, backwards-differencing
method to solve the coupled set of rate equations is fast and accurate
(see Bromm et al. 2002).

To achieve the dynamical range required to follow the evolution of the gas
from cosmological initial conditions down to the formation of high density
clumps, which are the immediate progenitors of Pop~III stars, and of the subseqent
supernova (SN) blast waves, we have implemented a resampling technique
within SPH (see Bromm \& Loeb 2003b for details). This method of refining
a coarser, parent simulation, and following the further collapse with increased
resolution by `splitting' the parent SPH particles into a number, $N_{\rm ref}$,
of child particles has already been successfully applied to problems in
Galactic star formation (Kitsionas \& Whitworth 2002).

\section{Simulations: Pre-Explosion Phase}
\subsection{Initial Conditions and Clump Formation}
We initialize our simulations at $z_{i}=100$ according to a standard
$\Lambda$CDM model with parameters $\Omega_{m}=1 - \Omega_{\Lambda}=0.3$, 
$\Omega_{B}=0.045$, $h=0.7$, and $\sigma_{8}=0.9$, close to the values 
measured by {\it WMAP} (Spergel et al. 2003).
Our computational box is periodic, has a comoving length of $100 h^{-1}$kpc,
and contains $128^{3}$ particles each in dark matter and gas. For the initial gas
temperature we adopt the value of 200~K. The fractional free-electron abundance
is initialized as $x_{e}=4.6\times 10^{-4}$, and the hydrogen molecule abundance
as $f_{{\rm H}_{2}}=2\times 10^{-6}$ (Anninos \& Norman 1996).

We follow the evolution until $z\simeq 20$, at which point the gas in one of
the halos has collapsed to a density $n\ga 10^{4}$cm$^{-3}$, thus violating
the requirement that the local Jeans mass be resolved everywhere: $M_{J} >
M_{\rm res}\simeq 500 M_{\sun}$ (e.g., Bate, Bonnell, \& Bromm 2003).
This first high-density clump to form in the simulation comprises a mass of
$\sim 10^{3} M_{\sun}$ within a radius of $\sim 1$~pc. The clump lies at the 
center of a virialized dark matter minihalo of total mass $\sim 10^{6} M_{\sun}$
and virial radius $\sim 150$~pc. If allowed to collapse further, the gas in the
clump would eventually form one very massive star (VMS), 
or a small multiple of them (see Abel et
al. 2002; Bromm et al. 2002). Here, we assume that the clump will
form a single VMS, producing ionizing radiation during its short lifetime
before finally exploding as a PISN. Next, we describe the radiative feedback that
precedes the explosion.

\subsection{Radiative Feedback}

To be able to simulate the propagation of the SN remnant into the
surrounding gas with sufficient resolution, we have selected a Cartesian
volume around the location of the high-density clump. The SPH particles
within this volume of physical length 1~kpc serve as the parent particles
for the resampling procedure, such that each parent spawns $N_{\rm ref}=16$
child particles (see Bromm \& Loeb 2003b for details). After refinement, there
are $\sim 5\times 10^{5}$ SPH particles in the dark matter minihalo that
hosts the PISN explosion, and the improved mass resolution is $M_{\rm res}\simeq
20 M_{\sun}$. The dark matter component, on the other hand, is not refined.

It is crucial to take into account the radiative feedback from the VMS on
the density structure in the minihalo. Otherwise, we would severely overestimate
the radiative losses in the early stages of the explosion. 
Recently, Bromm \& Loeb (2003c) have implemented
an approximate scheme to address this radiative feedback that allows one
to estimate the evolution of the central density field without
carrying out a fully self-consistent radiative transfer calculation.
We here briefly summarize this approach.

\begin{center} 
\psfig{file=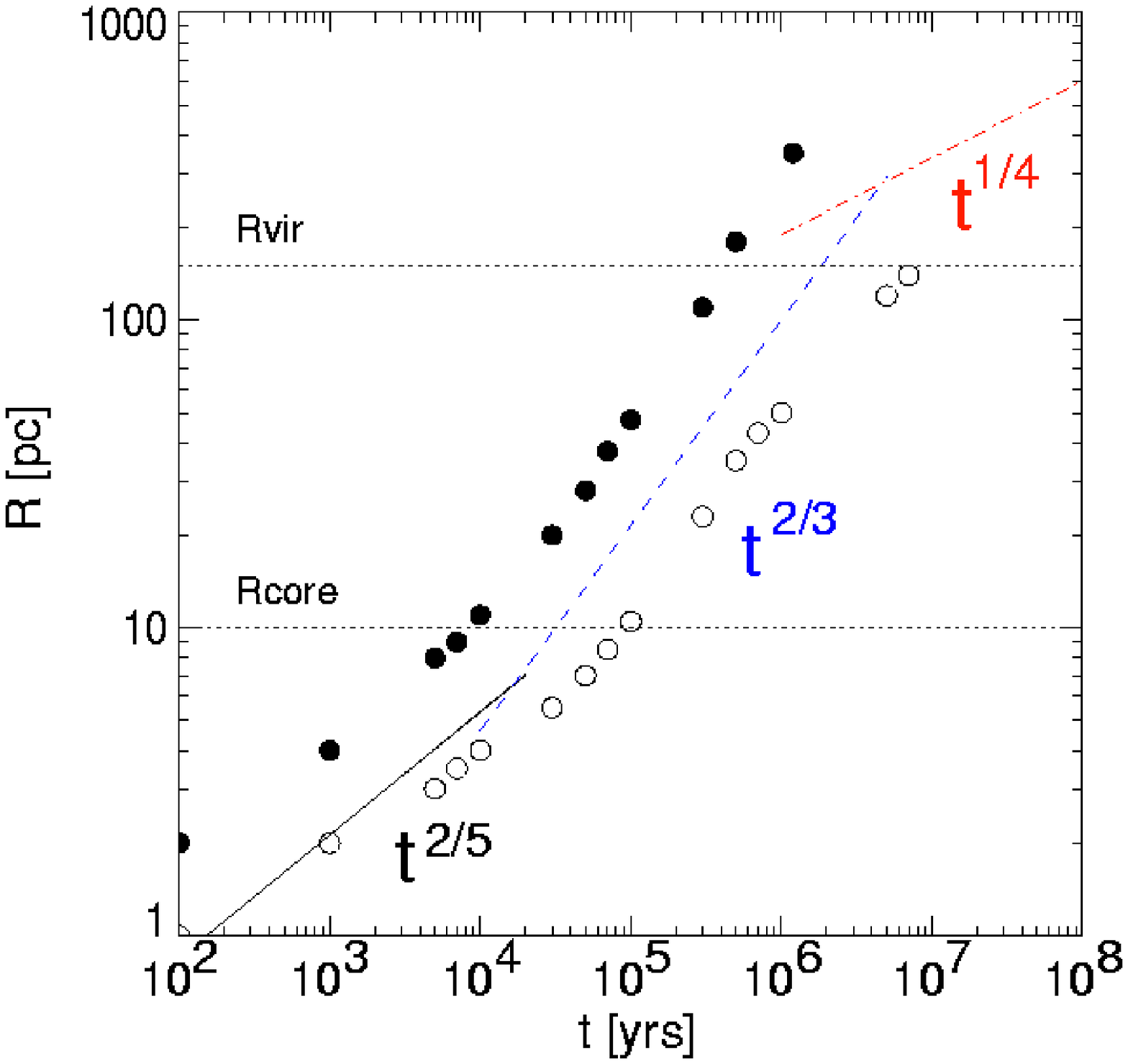,width=8.4cm,height=7.56cm}
\figcaption{
Evolution of the SN bubble. Shown is the radial position of the
dense shell (in pc) vs. time after the explosion (in yr).
{\it Filled symbols:} Simulation results for an explosion energy
$E_{\rm SN}=10^{53}$ergs.
{\it Open symbols:} Simulation results for an explosion energy
$E_{\rm SN}=10^{51}$ergs.
The lines show the analytical behavior of a SN blast wave. 
From early to late times: ({\it i}) Standard Sedov-Taylor
solution. This is applicable within $r<R_{\rm core}$ where the density
is roughly constant.
({\it ii}) Sedov-Taylor evolution in a $\rho \propto r^{-2}$ density profile.
This is the relevant case for $r>R_{\rm core}$. 
({\it iii}) Finally, at times $t \ga 10^{6}$yr, radiative losses become
important, and the SN remnant enters its snowplow phase.
\label{fig1}}
\end{center}

We assume that the gas within the minihalo is rapidly ionized on a timescale
that is short compared to the dynamical time in the halo. Specifically, we
initialize our feedback simulation, aimed at determining the hydrodynamic response
to the internal ionizing radiation, with a fully-developed \ion{H}{2} region
already in place. We choose a Str\"{o}mgren radius of $R_{\rm S}\simeq 200$~pc, and
assume that the ionizing radiation is unattenuated within $R_{\rm S}$ (Osterbrock 
1989). We now include photoheating terms in the entropy equation, such that
(in units of ergs s$^{-1}$ cm$^{-3}$): $\Gamma \propto r^{-2}\int {\rm d}\nu
f_{\nu}(R_{\ast})\sigma_{\nu}(\nu-\nu_{\rm th})/ \nu_{\rm th}$ for $r<R_{\rm S}$, 
and $\Gamma = 0$ otherwise.
Here, $f_{\nu}(R_{\ast})$ is the flux at the surface of a very massive Pop~III 
star (Bromm, Kudritzki, \& Loeb 2001b). We have
also included the relevant photoreactions in the chemical reaction 
network (see Haiman et al. 1996). We allow the photoheating due to the central
point source to go on for $\tau_{\rm MS}\simeq 2\times 10^{6}$yr, the approximate
main-sequence lifetime of a VMS. Under the influence of the Pop~III radiation
field, the gas inside the \ion{H}{2} region relaxes to a temperature of $T\simeq 3
\times 10^{4}$K, and the density within a central core of $R_{\rm core}\sim 20$~pc
reaches $n\sim 1$~cm$^{-3}$.

At the end of the radiative feedback phase, we switch off the central point source,
and initialize the SN explosion inside this photoheated density field.
The detailed density structure very close to the VMS, at radii $r\la 1$~pc, is
not reliably calculated by our approximate procedure. As we are here primarily
interested in the evolution of the SN shell at radii $\ga 10$~pc, this caveat
is not expected to compromise the main results in this study.

\section{Simulations: Post-Explosion Phase}

We carry out two explosion simulations. In each case, we assume that the
Pop~III star in the center of the minihalo explodes as a PISN. To span
the possible energy range, we consider explosion energies of
$E_{\rm SN}=10^{51}$~and $10^{53}$ergs, corresponding to stars with masses
$M_{\ast}\simeq 150$ and $250 M_{\sun}$ (Fryer et al. 2001).
We focus in particular on the
latter case, which marks
the upper mass limit of the PISN regime. 
We insert the explosion energy as thermal energy, distributing
it amongst $\sim 500$ SPH particles. These particles lie within
an initial radius of $\sim 2$~pc, such that the explosion is initialized
with conditions appropriate for the adiabatic Sedov-Taylor (ST) phase.
Since the PISN is completely disrupted without leaving a remnant behind, we
can trace the subsequent fate of the metals by using the SPH particles
that represent the stellar ejecta as markers.

\begin{center} 
\psfig{file=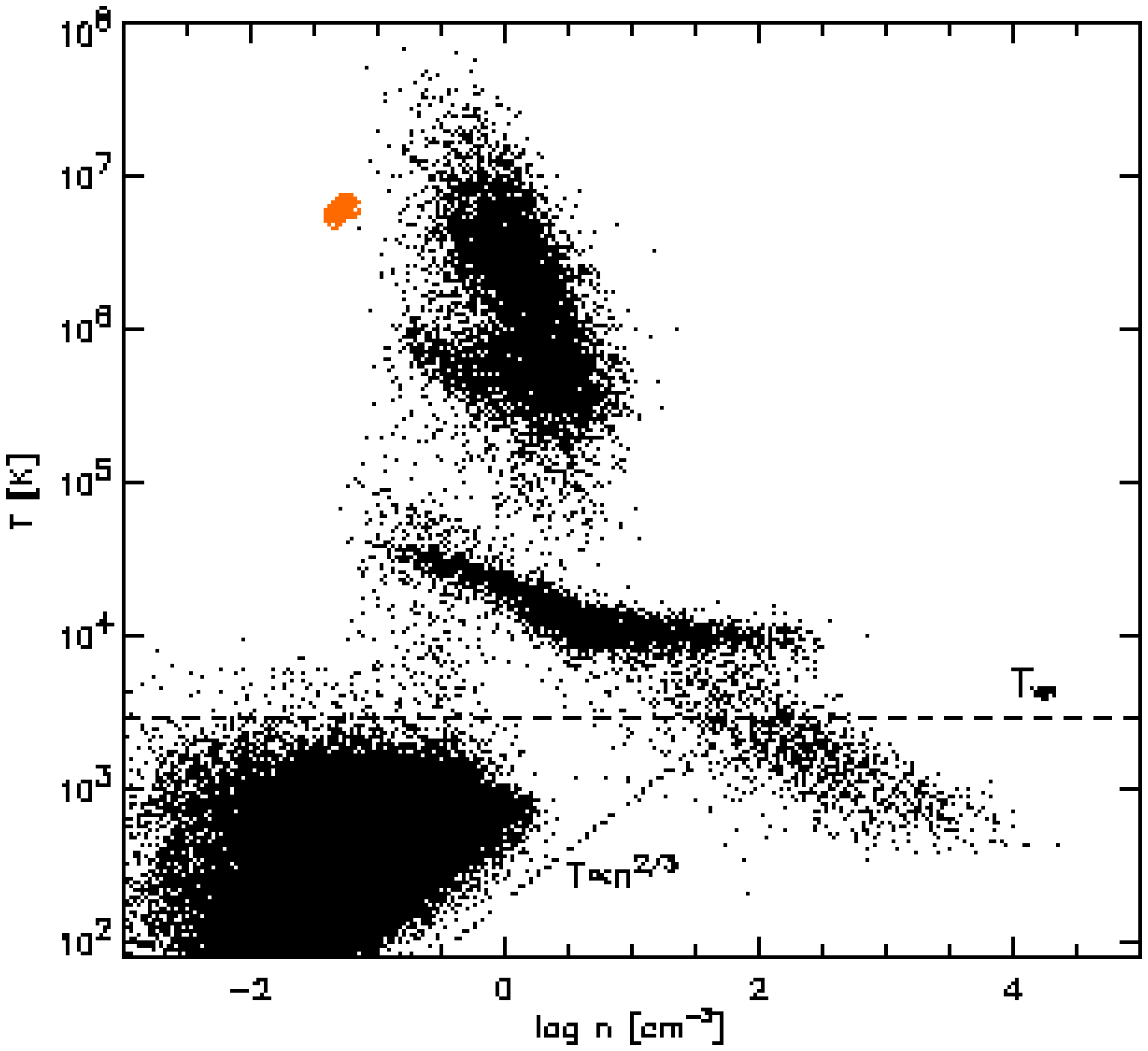,width=8.4cm,height=7.56cm}
\figcaption{
Thermodynamic properties of the shocked gas $\sim 10^5$~yr after the explosion.
Gas temperature (in K) vs. log $n$ (in cm$^{-3}$).
{\it Dashed line:} The virial temperature of the minihalo. It is evident that
$T_{\rm vir}\sim 3000$~K is much lower than the gas temperature in the minihalo.
Most of the gas will therefore be expelled from the halo on a 
dynamical timescale.
{\it Dotted line:} Adiabatic behavior. The gas in the outer reaches of the
minihalo, as well as in the general IGM, approximately follows this scaling.
{\it Red symbols:} Location of the gas that makes up the original
stellar ejecta. This gas is hot and diffuse, can consequently cool
only slowly, and therefore provides the pressure source that drives
the expansion of the bubble.
Notice that a fraction of the gas has been able to cool to $\sim 200$~K
due to the action of molecular hydrogen. 
\label{fig2}}
\end{center}

The expansion of the SN blast wave proceeds as shown in Figure~1, where
we compare our simulations with simple analytical calculations.
Initially, the blast wave evolves into a roughly uniform medium at radii
$r\la R_{\rm core}\sim 20$~pc, resulting in the usual ST scaling:
$R_{\rm sh}\propto t^{0.4}$. Once the blast wave reaches beyond the core, however,
it encounters an isothermal density profile in the remainder of the halo,
leading to the scaling (e.g., Ostriker \& McKee 1988):
$R_{\rm sh}\simeq 30{\rm \,pc} (E_{\rm SN}/10^{53}{\rm ergs})^{1/3}
(t/10^{5}{\rm yr})^{2/3}$. A few $10^{6}$yr after the explosion, radiative
(inverse Compton) losses become important, and the SN remnant enters its final,
snowplow phase. 
The significance of inverse Compton cooling distinguishes very high-$z$
SN explosions from those occuring in the present-day universe.
From Fig.~1, it is evident that the numerical simulations nicely reproduce 
the analytical expectation.

\begin{center} 
\psfig{file=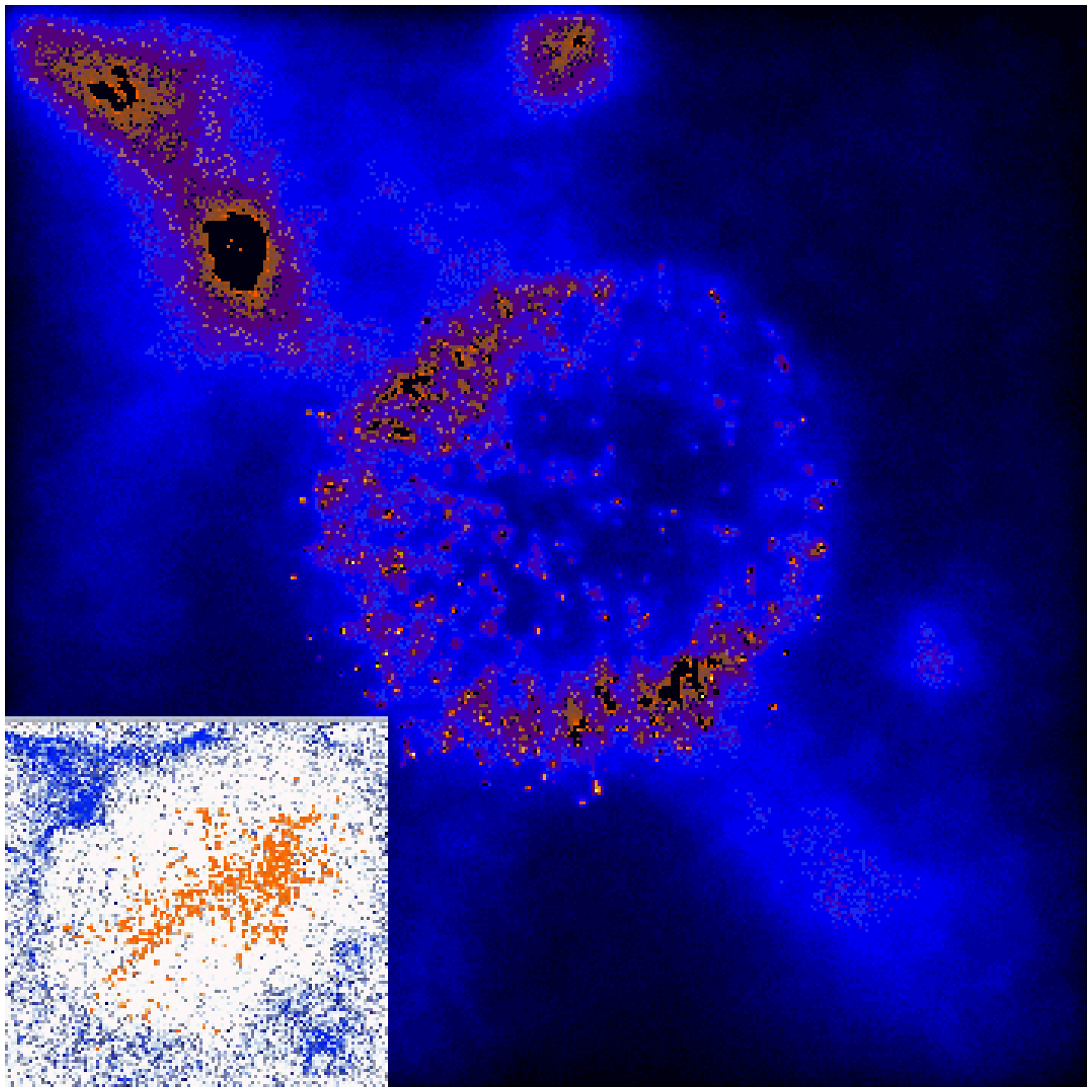,width=8.4cm,height=8.16cm}
\figcaption{
Situation $\sim 10^6$~yr after the explosion for the case with
$E_{\rm SN}=10^{53}$ergs. Shown is the projected gas density 
within a box of linear physical size 1 kpc.
The SN bubble has expanded to a radius of $\sim 200$~pc, having evacuated
most of the gas in the minihalo (with a virial radius of $\sim 150$~pc).
The dense shell has fragmented into numerous cloudlets.
{\it Inset:} Metal distribution after 3~Myr. The stellar ejecta ({\it red dots})
trace the metals, and are embedded in pristine, un-enriched gas ({\it blue dots}).
A large fraction ($\ga 90$\%) of the heavy elements has escaped the
minihalo, and filled a significant fraction of the hot bubble.
\label{fig3}}
\end{center}

After approximately $10^{5}$yr, when $t_{\rm cool} \la R_{\rm sh}/u_{\rm sh}$,
a dense shell begins to form at a radius $\sim 50$~pc. The thermodynamic
properties of the dense shell, the interior bubble, and the surrounding medium
are summarized in Figure~2. The shocked, swept-up gas exhibits three distinct
thermal phases, as can clearly be discerned in Fig.~2: the first at $T\ga 10^{5}$K,
the second at $T\sim 10^{4}$K, and the third at $T\la 10^{3}$K, with the latter
two corresponding to the dense shell. The cooling from very high temperatures 
to $\sim 10^{4}$K, and subsequently to $10^{2}$K, proceeds in non-equilibrium,
such that a fraction of free electrons persisits below $\sim 8,000$~K, allowing
the re-formation of H$_{2}$ (e.g., Shapiro \& Kang 1987;
Oh \& Haiman 2002). The molecular fraction
in the densest regions asymptotically assumes a `plateau' value
of $f_{{\rm H}_{2}}\sim 3\times 10^{-3}$. Inside the expanding SN remnant
lies a hot pressurized bubble (the red symbols in Fig.~2) that corresponds
to the original stellar ejecta. This bubble drives the outward motion, and its gas
is slowly cooled by adiabatic expansion and inverse Compton losses.
Initially, cooling due to adiabatic expansion dominates (see the distribution
of the red symbols in Fig.~2 as compared to the dotted line).

In Figure~3, we show the projected gas density $\sim 10^{6}$yr after the
explosion for the $E_{\rm SN}=10^{53}$ergs case. Two features are apparent:
Firstly, the blast wave has succeeded in completely disrupting the
minihalo, dispersing most of the halo gas into the IGM. Comparing $E_{\rm SN}$
with the binding energy of the minihalo, $E_{\rm grav}\sim 10^{51}$ergs,
such an outcome is clearly expected. It depends, however, on the inefficient
radiative processes that cool the hot bubble gas in the initial stages of
expansion. The evolution of the $E_{\rm SN}=10^{51}$ergs
case indeed proceeds very differently. The blast wave effectively stalls
at the virial radius, and the minihalo does not suffer complete disruption.

The second interesting feature in Fig.~3 is the vigorous fragmentation of
the dense shell into a large number of small cloudlets. 
The emergence of these cloudlets, corresponding to gas at $T\la
10^{3}$K, is not driven by gravity (e.g., Elmegreen 1994 and references
therein), but is instead due to a thermal instability triggered by the
onset of atomic cooling. 
The explosions simulated here do not result in the formation of lower-mass
stars, or Pop~II.5 stars in the terminology of Mackey et al. (2003).
This is consistent with the prediction
in Mackey et al. (2003) that Pop~II.5 stars can only form in dark matter halos
massive enough to be able to cool via atomic hydrogen (see also Salvaterra,
Ferrara, \& Schneider 2003).
To further test the possibility of SN triggered star formation in the high-redshift
universe, we 
will explore a range of cosmological
environments, corresponding to different halo masses and collapse redshifts,
in future work.

What is the fate of the metals that are produced in the PISN progenitor?
PISNe are predicted to have substantial metal yields, of the order
of $y=M_{Z}/M_{\ast}\sim 0.5$, and for the largest stellar masses in
the PISN range, most of $M_{Z}$ is made up of iron 
(Heger \& Woosley 2002). We find that in our $E_{\rm SN}=10^{53}$ergs 
simulation, $\sim 90$\% of the metals have escaped from the DM minihalo
$4\times 10^{6}$yr after the explosion. The metals have filled most
of the interior hot bubble (see inset of Fig.~3), and there has not yet been
sufficient time for them to significantly mix into the
dense surrounding shell. The eventual metal distribution is rather
flattened, with most metals ending up in the low-density voids.
A rough estimate of the resulting metallicity in the
surrounding, contaminated IGM can be obtained as follows: $Z=M_{Z}/M_{\rm gas}
\sim 100 M_{\sun}/10^{5}M_{\sun}\ga 10^{-2}Z_{\sun}$. This value is well above
the critical metallicity threshold, $Z_{\rm crit}\sim 10^{-3.5}Z_{\sun}$, such
that very massive stars can only form out of gas with $Z<Z_{\rm crit}$
(Omukai 2000; Bromm et al. 2001a; Schneider et al. 2002). We conclude
that Pop~III stars can continue to form in the pristine gas within the
neighboring halos, as the metals that are dispersed in the explosion 
do not reach them.

\section{Summary and Conclusions}

We have simulated the death of the first stars in minihalos at $z\simeq 20$,
assuming that they end their lives as PISNe.
For the highest possible explosion energies, $E_{\rm SN}=10^{53}$ergs,
we find that a large fraction ($\sim 90$\%) of the stellar metals are
dispersed into the surrounding IGM. 
The extent of the metal enriched region, with a physical size of $\sim 1$~kpc,
is comparable to the radius of the relic {\ion{H}{2}} region around the minihalo.
In a companion paper (Yoshida, Bromm, \& Hernquist 2003), we argue
that if Pop~III stars have led to an early partial reionization of the universe,
as may be required by the recent {\it WMAP} results (e.g., Wyithe \& Loeb 2003b),
this will have resulted in a nearly uniform enrichment of the universe to
a level $Z_{\rm min}\ga 10^{-4}Z_{\sun}$ already at $z\ga 15$.

For the most energetic PISNe, these metals are predominantly
in the form of iron, and the systems investigated in this {\it Letter}
could have been responsible for an early burst of Fe enrichment. On nucleosynthetic
grounds, such an initial episode of star formation has been inferred by
Qian \& Wasserburg (2002). Indeed, the `prompt inventory' postulated
by these authors could have arisen in the PISN explosions studied here.
Our simulations may also be relevant for explaining the recent observations
of quasars at $z\ga 6$, exhibiting surprisingly strong {\ion{Fe}{2}} lines  
(Freudling, Corbin, \& Korista 2003). These authors have interpreted
their results as hinting at an early nucleosynthetic contribution
from Type~Ia SNe, but it might be challenging to accomodate the required
long evolutionary timescales, first to form a white dwarf, and then
to sufficiently accrete from a binary companion to trigger the explosion.
In fact, if our picture is correct that normal Pop~II stars can only form
at $z\la 15$ (Mackey et al. 2003), the available time between $z\sim 15$ and
$\sim 6$ is only $\sim$0.7~Gyr, rendering the Type~Ia scenario problematic. A PISN
origin and the short timescales connected to it, however, could naturally
account for the high Fe/$\alpha$ ratios observed (see Heger \& Woosley 2002).

It would indeed be a remarkable feature of the universe, if the first stars
had endowed the IGM with such a widespread, near-universal level of pre-enrichment.

\acknowledgments

We are grateful to Volker Springel for making available to us a version
of GADGET.  This work has been supported in part by NSF grant AST
00-71019. The simulations were performed at the Center for Parallel 
Astrophysical Computing at the Harvard-Smithsonian Center for
Astrophysics.


\clearpage

\end{document}